\documentclass[twocolumn,showpacs,preprintnumbers,amsmath,amssymb]{revtex4}

\def\undersim#1{\setbox9\hbox{${#1}$}{#1}\kern-\wd9\lower
    2.5pt \hbox{\lower\dp9\hbox to \wd9{\hss $_\sim$\hss}}}
\usepackage{amssymb}
\usepackage{amsmath}
\usepackage{graphicx}

\def\undersim#1{\setbox9\hbox{${#1}$}{#1}\kern-\wd9\lower
    2.5pt \hbox{\lower\dp9\hbox to \wd9{\hss $_\sim$\hss}}}

\def\mk{{\mathbf k}}

\begin{document}

\title{Finite size of hardons and HBT interferometry for hydrodynamic
sources}

\author{Yong Zhang$^{1,\,2}$\footnote{zhy913@jsut.edu.cn}}
\author{Hong-Jie Yin$^3$}
\author{Weihua Wu$^{1}$}
\affiliation{\small$^1$School of Mathematics and Physics,
Jiangsu University of Technology, Changzhou, Jiangsu 213001, China\\
$^2$School of Science,Inner Mongolia University of Science $\&$ Technology, Baotou,Inner Mongolia Autonomous Region 014010, China\\
$^3$College of Mathematics and Physics, Bohai University, Jinzhou, Liaoning 121013, China\\
   }

\date{\today}

\begin{abstract}
Hadrons formed in heavy-ion collisions are not point-like objects, they cannot occupy
too close space-time points. When the two bosons are too close to each other,
their constituents start to mix and they cannot be considered as bosons subjected to Bose-Einstein
statistics, this effect is called excluded volume effect.
We study the volume effect on HBT for the sources with various sizes.
The effect on HBT was shown in out, side and long directions, and it is more obvious
for the source with a narrow space-time distribution.
The correlation functions for high transverse momenta are more suppressed by
the volume effect. Hence the incoherence parameter may be more suppressed by the volume effect for high transverse momenta in small collision systems.


\end{abstract}

\maketitle

\section{Introduction}
Hanbury-Brown-Twiss (HBT) interferometry has become an important tool for detecting the
space-time structure of the particle emitting sources formed in heavy-ion collisions
\cite{Won94,Wie99,Wei00,Lis05}. The correlation functions of two identical bosons were
defined as the ratio of the two-particle momentum spectrum $N(\mk_1,\mk_2)$ of identical
bosons to the product of the two single boson momentum spectra $N(\mk_1)$$N(\mk_2)$, and
the theoretical results of HBT are always above 1. However, the observed data show
that the two-pion Bose-Einstein correlations function takes values below unity \cite{L3,LEP2,RHIC11,CMS11}.

Hadrons formed in heavy-ion collisions are not point-like objects,
they cannot get too close to each other, this effect is called excluded volume effect \cite{Hagedorn_PLB,Yen_prc,Rischke_zpc}. When the two bosons are too close to each other,
their constituents start to mix and they cannot be considered as bosons subjected to Bose-Einstein
statistics, so the HBT experiment cannot see two bosons which are too close in space-time \cite{Bialas13,Bialas15,Bialas16}. In Ref. \cite{Bialas13}, A. Bialas $et$ $al$. first study
the excluded volume effect on HBT. Then the full Bose-Einstein correlation functions was shown
in three directions (out,side,long) for pp collisions at 7 TeV \cite{Bialas15}, using the blast-wave model \cite{Blast1,Blast2,Blast3}. The excluded volume effect on HBT relates to the size of hadrons and
the scale of sources.

In this paper, we study the excluded volume effect on HBT for the
sources with various sizes, using the ideal relativistic hydrodynamics in $2+1$ dimensions to describe the transverse
expansion of sources with zero net baryon density and combine the Bjorken boost-invariant
hypothesis \cite{Bjo83} for the source longitudinal evolution. In the calculations, we use the
equation of state of s95p-PCE, which combines the hadron resonance gas at
low temperatures and the lattice QCD results at high temperatures \cite{She10}.
We assume that the system reaches the static local equilibrium at $\tau_0=0.6$
fm/$c$ after the collision and take the initial energy density distribution
in the transverse plane as the Gaussian distribution,
\begin{equation}
\epsilon = \frac{\epsilon_0}{2{\pi}R_{x}R_{y}}\exp[-x^2/(2R_x^2)-y^2/(2R_y^2)],
\end{equation}
where $\epsilon_0$ and $R_i~(i=x,y)$ are the parameters of the initial source
energy density and radii.

The rest of this paper is organized as follows. In Sec. II, we present the calculating formula of
two-pion correlation function for hydrodynamic sources. In Sec. III, we show the excluded volume
effect on pion HBT. Finally, a summary and conclusion of this paper are given in Sec. IV.

\section{Calculations of HBT for hydrodynamic sources}
The Bose-Einstein correlation function was usually defined as \cite{Wie99,Wei00,Lis05}
\begin{eqnarray}
&&\hspace*{-8mm}C(\mk_1,\mk_2) =\frac{N(\mk_1,\mk_2)}{N(\mk_1)N(\mk_2)} \\\nonumber
&&\hspace*{7.5mm}= 1+\frac{|\int d^{4}rS(r,K)e^{iq\cdot r}|^{2}}{\int d^{4}r_{1}S(r_{1},k_1)d^{4}r_2S(r_2,k_2)},
\end{eqnarray}
where $S(r,k)$ is the emission function. $k_1$, $r_1$ and $k_2$, $r_2$ are momenta and
 positions of the particles. $q = k_1-k_2$, $K = (k_1+k_2)/2$ are the relative momentum and
pair momentum of two identical particles, respectively. With the ``smooth assumption'' $S(r,(k_1+k_2)/2)\approx S(r,k_1)\,\approx S(r,k_2)$,
the Bose-Einstein correlation function can be rewritten as \cite{Lis05,QHZHang1997,Pratt1997,Anc1998}
\begin{eqnarray}\label{hbtfunc}
&&\hspace*{-7mm}C_0(\mk_1,\mk_2) = 1+ \\\nonumber
&&\hspace*{7mm}\frac{\int d^{4}r_1d^{4}r_2S(r_1,{k}_1)S(r_2,k_2)\cos(q\cdot(r_1-r_2))}{\int d^{4}r_1S(r_1,{k}_1)d^{4}r_2S(r_2,k_2)}.
\end{eqnarray}

Considering the excluded volume effect, the HBT correlation function becomes \cite{Bialas13,Bialas15,Bialas16}
\begin{equation}\label{hbtfunc1}
C_{v}(\mk_1,\mk_2) =  C_0(\mk_1,\mk_2)- C_{c}(\mk_1,\mk_2),
\end{equation}
the corrected part of the correlation function $C_{c}(\mk_1,\mk_2)$ was defined as \cite{Bialas15}
\begin{equation}\label{hbtcorr}
C_{c}(\mk_1,\mk_2) =  C_{c}^{(0)}(\mk_1,\mk_2)+ C_{c}^{(q)}(\mk_1,\mk_2),
\end{equation}
\begin{equation}
C_{c}^{(0)}(\mk_1,\mk_2)=\frac{\int d^{4}r_1d^{4}r_2S(r_1,{k}_1)S(r_2,k_2)D_{12}}{\int d^{4}r_1S(r_1,{k}_1)d^{4}r_2S(r_2,k_2)},
\end{equation}
\begin{eqnarray}
&&\hspace*{-1mm}C_{c}^{(q)}(\mk_1,\mk_2)= \\\nonumber
&&\hspace*{3mm}\frac{\int d^{4}r_1d^{4}r_2S(r_1,{k}_1)S(r_2,k_2)\cos(q\cdot(r_1-r_2))D_{12}}{\int d^{4}r_1S(r_1,{k}_1)d^{4}r_2S(r_2,k_2)},
\end{eqnarray}

where $D_{12}$ is the cut-off function. For hydrodynamic
sources, we treat the cut-off function as
\begin{equation}
D_{12}=e^{-d^2(r_1,r_2)/{\Delta}r^2-d^2(t_1,t_2)/{\Delta}t^2},
\end{equation}
where $d^2(r_1,r_2)=(x_1-x_2)^2+(y_1-y_2)^2+(z_1-z_2)^2$ and $d^2(t_1,t_2)=(t_1-t_2)^2$
are the total distance and time difference between two particles, respectively.
The cut-off distance  ${\Delta}r\approx 2 r_V$  (where $r_V$ is
the radius of the ``excluded volume''~\cite{Hagedorn_PLB,Yen_prc,Rischke_zpc,Gor1981} occupied by one pion)
was taken to be 1 fm \cite{Bialas15}. The cut-off time difference $\Delta t$ is newly introduced in this
study, and it represents the time required for two particles to leave the overlap region.
The cut-off time difference $\Delta t$ is treated as a constant parameter, and it is
taken to be o.2 fm/c in this paper. It can be seen
from the Eq.\,(4)\,-\,(7), the volume effect on the correlation function is sensitive to the space-time
distribution of the pion source, which will be larger for a narrow space-time
distribution of the pion source.

On the freeze-out surface of hydrodynamic sources, the pion emission
function is defined by the Cooper-Frye integral \cite{Cooper1974,Sch1992,Heinz-PLB1994,Heinz-PRC2015}
\begin{equation}\label{emission}
S(r,p) = \frac{1}{(2\pi)^3}\int_{\Sigma}\frac{p^{\mu}  d^3 \sigma_{\mu}(r^{\prime})\, \delta^4 (r-r^{\prime})}{\exp[p^{\mu}u_{\mu}(r^{\prime})/T]-1}.
\end{equation}
Where $u_{\mu}(r)$ is the flow velocity profiles along the freeze-out surface $\Sigma$, and $d^3\sigma_{\mu}(r)$ is the outward pointing normal vector on $\Sigma$ at point $r$. The pion freeze-out temperature $T$ is taken to be 130 MeV \cite{Hu15} in this paper.

\section{Results}

\subsection{Pion emission source}
\begin{figure}[htb]
\vspace*{2mm}
\hspace*{7mm}
\includegraphics[scale=0.55]{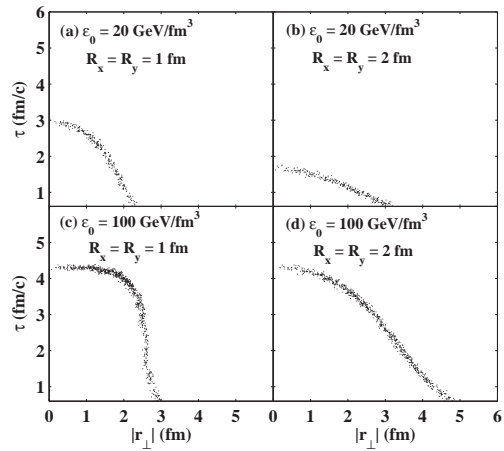}
\vspace*{-4mm}
\caption{Distributions of the freeze-out points of pion in the $z=0$ plane for various initial conditions.}
\label{rt}
\end{figure}

In Fig. \ref{rt}, we show the space-tmie distributions of the freeze-out points of pion in the $z=0$ plane for various initial conditions. Where the $\tau$ and the $r_{\perp}$ are the time and the transverse coordinate, respectively, in the $z=0$ plane. For a fixed $\epsilon_0$, the spatial distribution of the pion source for $R_x = R_y =$ 2 fm
is wider than for $R_x = R_y =$ 1 fm, and the emission time of pion for $R_x = R_y =$ 2 fm is earlier than $R_x = R_y =$ 1 fm.

\begin{table}[hbt]
\caption{Spatio-temporal properties of pion emission source. Where the initial conditions (a)-(d) are the same as in Fig. \ref{rt}.}
\begin{tabular}{cccccc}
\hline\hline
Initial condition~&~$\bar{\tau}$\,(fm/\emph{c})~&~$\overline{|r_{\bot}|}\,$(fm)~&
~${\sigma_{\tau}}$\,(fm/\emph{c})~&~${\sigma_{r}}$\,(fm)\\
\hline
(a)~&~2.07~&~1.37&~0.67&~0.53\\
(b)~&~1.14~&~1.94&~0.30&~0.76\\
(c)~&~3.60~&~1.85&~0.94&~0.65\\
(d)~&~2.82~&~2.66&~1.09&~1.10\\
\hline\hline
\end{tabular}
\label{Tab-1}
\end{table}

In table \ref{Tab-1}, we show some spatio-temporal properties of pion emission source. Where the initial conditions (a)-(d) are the same as in Fig. \ref{rt}. $\bar{{\tau}}$ and $\overline{|r_{\bot}|}$ are the average time and the average transverse radius, respectively, of pion freeze-out points in the $z=0$ plane.
$\sigma_{\tau}$ and $\sigma_{r}$
are the standard deviation of the freeze-out time and transverse radius:
\begin{eqnarray}
\label{Sigmat}
\sigma_{\tau} = \sqrt{\frac{1}{N}\sum_{i=1}^{N}(\,{\tau}_{i}-\bar{{\tau}}\,)^{\,2}},
\end{eqnarray}
\begin{eqnarray}
\label{Sigmat}
\sigma_{r} = \sqrt{\frac{1}{N}\sum_{i=1}^{N}(\,{|{r_\bot}|_i}-\overline{|r_{\bot}|}\,)^{\,2}}.
\end{eqnarray}
Where $N$ is the total number of the freeze-out points. ${\tau}_{i}$ and ${|{r_\bot}|_i}$
are the space-time coordinate of the freeze-out point denoted by $i$. The ${\sigma_{\tau}}$ and ${\sigma_{r}}$
can represent the width of the space-tmie distributions of the pion freeze-out points in the $z=0$ plane. For fixed initial radius, the width increase with the increasing initial energy density $\epsilon_0$.

\subsection{HBT results}
\begin{figure}[htbp]
\vspace*{7mm}
\hspace*{0mm}
\includegraphics[scale=0.8]{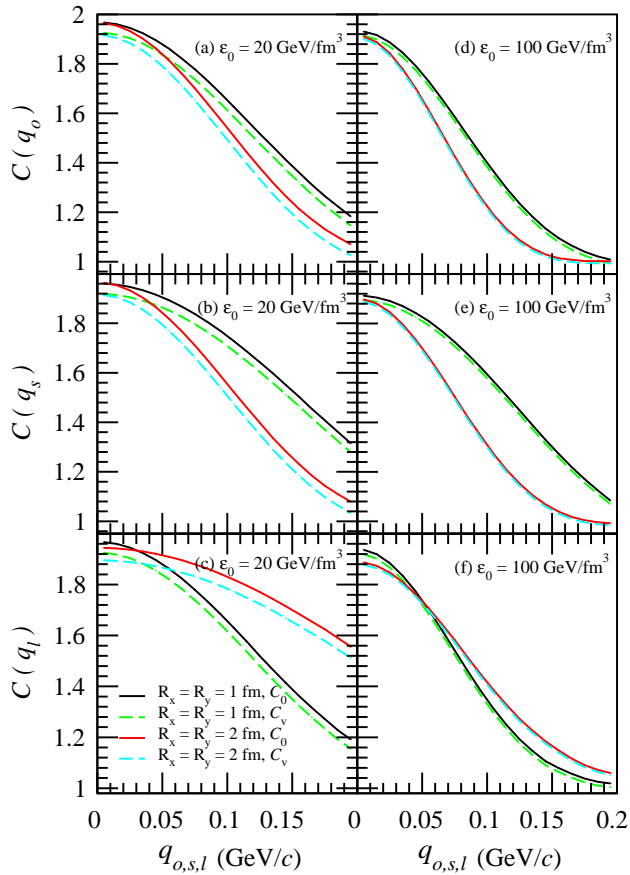}
\vspace*{-3mm}
\caption{Correlation function for out (top panel), side (middle panel) and long direction (bottom panel) in the interval 0.2 GeV/$c$ $\leq$ $K_T$ $\leq$ 0.4 GeV/$c$. For a fixed direction, the relative momentum $q$ for other two directions are less than 40 MeV/$c$. Where $C_0$ is the correlation function without the volume effect, and
the $C_v$ is the correlation function with the volume effect.}
\label{c2}
\end{figure}

In Fig. \ref{c2}, we show the correlation function for out (top panel), side (middle panel) and long direction (bottom panel) in the interval 0.2 GeV/$c$ $\leq$ $K_T$ $\leq$ 0.4 GeV/$c$. For a fixed direction, the relative momentum $q$ for other two directions are less than 40 MeV/$c$. In the calculations, we take the spatial rapidity $\eta_s$ in the region $(\,-1,\,1\,)$. Here, $C_0$ is the correlation function without the volume effect, and the $C_v$ is the correlation function with the volume effect.

\begin{figure}[htbp]
\vspace*{7mm}
\hspace*{0mm}
\includegraphics[scale=0.8]{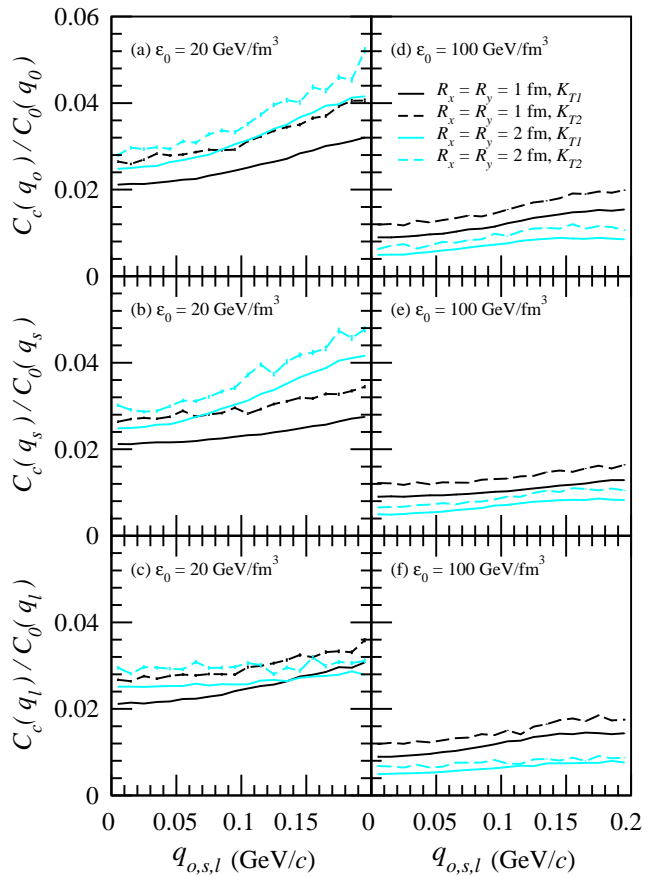}
\vspace*{-3mm}
\caption{The ratio of $C_c$ to $C_0$ for out (top panel), side (middle panel) and long direction (bottom panel).
The results for the marks of $K_{T1}$ and $K_{T2}$ are calculated in the momentum regions
0.2 GeV/$c$ $\leq$ $K_T$ $\leq$ 0.4 GeV/$c$ and 0.4 GeV/$c$ $\leq$ $K_T$ $\leq$ 0.6 GeV/$c$.
Solid lines depict the results for $K_{T1}$, and the dashed lines depict the results for $K_{T2}$.}
\label{ccor}
\end{figure}
In Fig. \ref{ccor}, we show the ratio of $C_c$ to $C_0$ for out (top panel), side (middle panel) and long direction (bottom panel). Where the $C_c$ is the corrected part of the correlation function.
The results for the marks of $K_{T1}$ and $K_{T2}$ are calculated in the momentum regions 0.2 GeV/$c$ $\leq$ $K_T$ $\leq$ 0.4 GeV/$c$ and 0.4 GeV/$c$ $\leq$ $K_T$ $\leq$ 0.6 GeV/$c$.
Solid lines depict the results for $K_{T1}$, and the dashed lines depict the results for $K_{T2}$.
Here, the ratio can quantitatively represent the volume effect on the correlation function.
The Fig. \ref{c2} and Fig. \ref{ccor} present the following four phenomena:

{\bfseries I.} The correlation functions are more suppressed by the volume effect for $\epsilon_0$ = 20 GeV/fm$^3$ than for $\epsilon_0$ = 100 GeV/fm$^3$.

{\bfseries II.} For $\epsilon_0$ = 20 GeV/fm$^3$, the correlation functions are more suppressed for $R_x = R_y =$ 2 fm than $R_x = R_y =$ 1 fm (see Fig. \ref{c2} (a)-(c)).

{\bfseries III.} For $\epsilon_0$ = 100 GeV/fm$^3$, the correlation functions are more suppressed for $R_x = R_y =$ 1 fm than $R_x = R_y =$ 2 fm (see Fig. \ref{c2} (d)-(f)).

{\bfseries IV.} For a fixed initial condition, the volume effect on the correlation function for high momentum is greater than for lower momentum.

With the Bjorken boost-invariant hypothesis \cite{Bjo83}, the pion emission time and the
longitudinal coordinate were expressed as

\begin{eqnarray}
\label{Bj-z}
t = \tau \cosh{\eta_{s}},
\end{eqnarray}
 \begin{eqnarray}
\label{Bj-t}
z = \tau \sinh{\eta_{s}},
\end{eqnarray}
for a fixed spatial rapidity $\eta_s$, the pion emission time $t$ and the
longitudinal coordinate are proportional to the emission time in the $z=0$ plane.
And the width of the time $t$ and the longitudinal coordinate increase with the increasing
spatial rapidity $\eta_s$ for a fixed distribution of the $\tau$.

\begin{figure}[htbp]
\vspace*{1mm}
\hspace*{-2mm}
\includegraphics[scale=0.88]{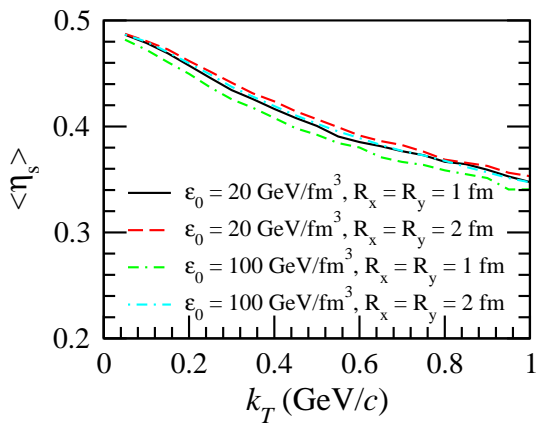}
\vspace*{-3mm}
\caption{The average emission spatial rapidity of the pion for different transverse momenta.}
\label{etas}
\end{figure}

In Fig. \ref{etas}, we show the average emission spatial rapidity of the pion for different transverse momenta.
For a fixed transverse momentum, the average emission spatial rapidity of the pion varies little
under different initial conditions. Thus, the main factor for phenomenon {\bfseries I.} is that
the width of the space-tmie distributions of the pion source in the $z=0$ plane for $\epsilon_0$ = 20 GeV/fm$^3$ are narrower than for $\epsilon_0$ = 100 GeV/fm$^3$ (see table \ref{Tab-1}).
For $\epsilon_0$ = 20 GeV/fm$^3$, the transverse spatial width parameter, $\sigma_r$,
of the pion source for $R_x = R_y =$ 1 fm is a little smaller than for $R_x = R_y =$ 2 fm. However, the temporal width parameter, $\sigma_{\tau}$, of the pion source for $R_x = R_y =$ 1 fm is much greater than for $R_x = R_y =$ 2 fm. Thus, the width of the time and the longitudinal space for $R_x = R_y =$ 2 fm are much narrower than for $R_x = R_y =$ 1 fm. And it is the main factor for phenomenon {\bfseries II.}.
For $\epsilon_0$ = 100 GeV/fm$^3$, the width of the space-tmie distributions of the pion source in the $z=0$ plane for $R_x = R_y =$ 1 fm are narrower than $R_x = R_y =$ 2 fm (see table \ref{Tab-1}). And it is the main factor for phenomenon {\bfseries III.}

\begin{figure}[htbp]
\vspace*{5mm}
\hspace*{0mm}
\includegraphics[scale=0.88]{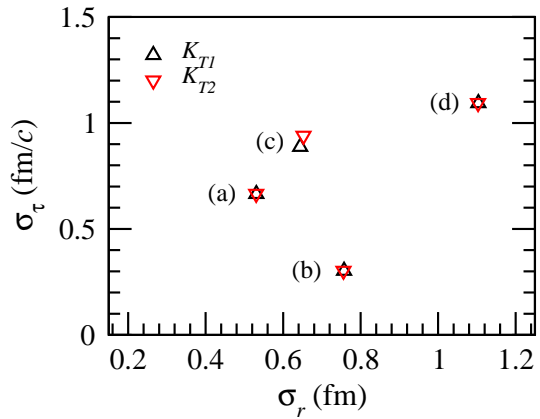}
\vspace*{-1mm}
\caption{The width parameters of time and space of the pion source in the $z=0$ plane for various initial conditions.
The marks (a) - (d) represent the initial conditions are as the same as in Fig. \ref{c2}.}
\label{sgm}
\end{figure}

In Fig. \ref{sgm}, we show the width parameters of time and space of the pion source in the $z=0$ plane for various initial conditions. The marks (a) - (d) represent the initial conditions are as the same as in Fig. \ref{c2}.
For a fixed initial condition, the width parameters of time and space of the pion source in the $z=0$ plane for different momenta are almost the same. The average emission spatial rapidity of the pion decreases with the increasing transverse momentum (see Fig. \ref{etas}). Thus, the width of the time and the longitudinal space
are narrower for large transverse momentum, and it leads to a greater volume effect on correlation function of the
large transverse momentum (phenomenon {\bfseries IV.}).

For a fixed initial condition, the correlation functions for high transverse momenta are more suppressed by
the volume effect. Hence the incoherence parameter $\lambda$ \cite{Won94,Wie99,Wei00,Lis05} may be more suppressed by the volume effect for high transverse momenta. Theoretically the incoherence parameter $\lambda$ can be less than unity due to partial coherence of strong interaction, long-lived resonance decays and the non-Gaussian form of the correlation function \cite{Wie99,Lis05,STAR04,STAR05,YGMa}. Since for increasing transverse momenta the resonance contributions
decrease, the $\lambda$-parameter increases with transverse momenta \cite{STAR05,Schlei,Bolz,Tcs}.
For small collision systems, the volume effect is obvious, and it may lead to a small slope of the $\lambda$-parameter
with respect to the transverse momentum.

\section{Summary}
Hadrons formed in heavy-ion collisions are not point-like objects, they cannot occupy
too close space-time points. When the two bosons are too close to each other,
their constituents start to mix and they cannot be considered as bosons subjected to Bose-Einstein
statistics, this effect is called excluded volume effect \cite{Hagedorn_PLB,Yen_prc,Rischke_zpc}.

In this paper, we study the volume effect on HBT for the sources with various sizes.
The effect on HBT was shown in out, side and long directions, and it is more obvious
for the source with a narrow space-time distribution.
For a fixed initial condition, the correlation functions for high transverse momenta are more suppressed by
the volume effect. And the incoherence parameter may be more suppressed by the volume effect for high transverse momenta. Hence it may lead to a small slope of the $\lambda$-parameter
with respect to the transverse momentum for small collision systems.

\begin{acknowledgments}
This research was supported by the National Natural Science Foundation
of China under Grant No. 11647166, the Natural Science Foundation of Inner Mongolia
under Grant No. 2017BS0104, Changzhou Science and Technology Bureau CJ20180054, and the Foundation of Jiangsu University of Technology
under Grant No. KYY17028, KYY18048.
\end{acknowledgments}

\end{document}